 \documentclass[10pt, a4paper]{article}
 \usepackage{graphicx}                    
 \usepackage{color}                       
 \usepackage{url}                         

\begin{document}



Title: Rotation of the photospheric magnetic field through solar cycles 21, 22, 23.

Authors: E. A. Gavryuseva (Institute for Nuclear Research RAS)

Comments: 8 pages, 3 Postscript figures

 \begin{abstract}
   Rotation of the large scale solar magnetic field  has a great importance 
 for the understanding of solar dynamic, 
 for the searh of longitidinal structure and  
 for the study of solar-terrestrial relations.  
    30-year long observations taken at the Wilcox Solar Observatory (USA) in 21-23 cycles  
    have been analyzed carefully to deduce  magnetic field
    rotation rate at different latitudes in both hemispheres
    and its variability in time. 
  
   The WSO data appear to indicate that 
   additionally to the differential rotation along the latitudes 
   there are running waves of fast rotation of the magnetic field.
   These torsional waves are running from the poles to the equator with a period of 11 years.
   
    The rotation of the magnetic field (RMF) is almost rigid at latitudes
    above 55 degrees in both hemispheres.
   The rotation rate in the sub-polar regions is slower
   when the magnetic field is strong there
   (during minima of solar activity),
   and faster when the magnetic field changes polarity
   (during maxima  of solar activity).
 \end{abstract}
 \vspace{1pc}
        \section{Introduction}
    The characteristics of the solar magnetic field and 
    their variability have been studied over the years by 
    many authors (see Gavryuseva (2018) and references there).
    The rotation of the magnetic field 
    is extremely important for the solar dynamo theory.
    
    This paper is focused on the study of the rotation 
    of the large scale photospheric magnetic field (SMF)
    and its variability through solar activity cycles.
    Rotation rate of solar plasma was measured over many years. 
    Sunspot spectroscopic observations provide the information 
    on the mid and low latitudes.
        
 
  The  Wilcox Solar Observatory  (WSO) data
  from http://wso.stanford.edu/ have been used because they provide
  the sets of the large scale photospheric  solar magnetic field (SMF)
  observations at different latitudes. 
  The rotational grid of the available data is made of 30
  equal steps  in  sine of latitude $\theta$ (sin($\theta$))
  where $\theta\pm39.3$, $\pm34.5$,  $\pm30.0$,
  $\pm25.7$,  $\pm21.5$, $\pm17.5$,  $\pm13.5$,  $\pm9.6$,  $\pm5.7$,
  $\pm1.9$ degrees, and of 5 degree steps in heliographic longitude.
  The observations were taken  by the WSO's Babcock solar magnetograph
  using the Zeeman splitting of the 525.02 nm Fe I spectral line 
  (Scherrer et al., 1977; Hoeksema, 1984).
  The resuls of the study of the rotational rate of the SMF data sets 
  since May 27, 1976 over 21, 22, 23 cycles are presented in this paper.

 \section{Rotation of the Solar Magnetic Field Through the Cycles}
   We have used two independent methods to evaluate the period of the 
   rotation of the solar magnetic field:
   spectral analysis (fast Fourier transform -- FFT) 
   and autocorrelation.
   The  periods of the differential rotation
   at each latitude were deduced for the
   whole  29 years long  data sets
   and for  sub-sets of shorter duration
   with step of 1 CR.
   The differential rotational period was also  estimated as a mean
   of the rotational periods calculated for the shorter series
   at each latitude.

    The results obtained  by the FFT method for the sub-sets
    of 40 CR (about 3 years) long
    are presented in Fig. 1.
    On the upper plot the sideral period of the magnetic field rotation
    as a function of time and latitude is shown.
    Blue (yellow) colors correspond to the  shorter (longer) periods.
    Contours correspond to the periods
    of 27, 28, 29, 30 and 31  days.
  The well known differential rotation appears also
  for the large scale solar magnetic field.
  Additionally in the sub-polar zones
  there is a clearly visible decrease of the rotational rate
  in 1985 and 1994  during solar activity minima
  and an increase of the rotation rate
  approximately in 1990 and in 1991 after
  the polarity inversion.
  This happens with the 11-year periodicity 
  (Gavryuseva, 2005, 2006, 2006d, 2008a,b).
  This result provides the panoramic understanding of the variability
  of the magnetic field rotatoion
  and complete the earlier studies performed by
  Gilman and Howard (1984),   Stenflo  (1977, 1990),
  Obridko and Shelting  (2001).
  An attempt of modeling  11-year variations in sub-polar regions
  was done by Tikhomolov (2001).
      
In Fig. 2 the SMF mean sideral  rotational period deduced from the full sets
of 29 year long (composed of 27721 points)
are plotted as a function of latitude deduced by auto-correlation method (continuous line)
as well as by the FFT method (dotted line) and as a mean  periods
deduced by the FFT method for 3-years long subsets (dashed-dotted line) with corresponding error bars.
There is  a 0.5--0.7-day decrease of the period at latitudes
higher than 56-60 degrees   in both hemispheres,
corresponding to the 1.7--2.3\% range.
The accuracy of the autocorrelation method for the full data sets
is limited by the longitudinal resolution of 5 degrees,
and its accuracy is  equal to 1.3\% at most.
 This result coincides with the latitudinal dependence of the
 rotation rate calculated  by the FFT method for the full data sets,
 and with the  rotation rate  calculated by both methods 
 as the mean of the rotation rates
 corresponding to the  shorter subsets.
 The accuracy of the mean rotation rate is at least 10 times better.

  The sideral SMF rotational period agrees with the results
  of the spectroscopic measurements of the solar rotation  in the
  interval of latitudes $\theta$ from -40 to 40  degrees
    (Howard and Harvey,    1970;
     Howard et al.,        1991),
  see also, for example,
    (Stenflo,              1974;
     Godoli and Mazzuconi, 1979, 1983;
     LaBonte and Howard,   1981, 1982b;
     Snodgrass,            1983;
     Howard et al.,        1984;
     Bumba and  Heina,     1987;
     Ulrich et al.,        1988;
     Snodgrass and Ulrich, 1990;
     Beck,                 1999;
     Ivanov et al.,        2001;
     Ossendrijver,         2003).

  On the contrary at the latitudes between 40 and 55 degrees
  the  SMF rotates faster than  other tracers and the
  photospheric plasma. This result  of Gavryuseva (2006, 2006d, 2008)
  and Gavryuseva \& Godoli (2006) well agrees
  with the latitudinal dependence of the rotational period
  of magnetic field data from the Mount Wilson and 
  Kitt Peak National Observatories taken in 1959-1985,
  reported by  Stenflo, (1989).
  It was noted also by  Obridko and Schelting  (2001)
  that the solar magnetic field
  rotates more rigid at high latitudes.
  The decrease of the SMF rotational period
  at latitudes above 55-60 degrees
 (deduced   from the 25-30 years long data sets of the SMF)
  has never been found
  for other tracers   or in spectroscopic measurements of
  shorter duration (Beck, 1999).
  This interesting result can be explained by
  the replenishing of surface magnetic field
  'over a time scale of weeks by new flux emitted from the source 
  region, which is probably near the bottom of the convection zone'   (Stenflo,  1989)
  (the rotation of the plasma at the bottom is faster than at the surface in high latitudes).
  The rigid SMF's rotation at high latitudes can explain the fact of slow rigid rotation of coronal holes.

  The radial and latitudinal dependence of the rotational rate
  was obtained by helioseismological methods, see for example,
      (Rhodes et al., 1990;
       Howe et al.,   2000;
       Gavryuseva et al., 2000;
       Di Mauro,      2003).
    The higher rotation rate at the latitudes above the 55 degrees
    corresponds to the rate of the rotation at a deeper layer.
   This  is an evidence that the solar magnetic field rotation
   at  latitudes above  55 degrees follows the rotation of the
   deeper layers where it is originated from.

   Such latitudinal dependence of the rotation
   would also support the theoretical model of
    Snodgrass  (1986, 1987a) and    Wilson  (1988)
  of the solar cycle based on the existence of
  a system of latitude- and
  time-dependent toroidal convective rolls in which the rotation rate
  is alternately faster or slower
  than the time averaged differential rotation at that latitude
  (Rabin et al., 1991).
  This model was suggested to describe the observed
  torsional oscillations.

  On the bottom plot of Fig. 1 the mean North-South deviation of the
  rotational period, obtained by the  FFT method for the 3 years long
  sub-sets, from the mean period over 1976-2004 years
  for each latitude $\theta$
  is shown as a function of time and  latitude.
  This interval corresponds to the cycles No 21 and No 22
  (two complete cycles have  been chosen
  in order to avoid possible influence of the variability through  cycles).
  Blue (yellow) colors correspond to negative (positive) deviation.
  Contours correspond to the deviations of $0, \pm0.5, \pm1.0$  days.

    Torsional waves firstly discovered by Howard and LaBonte
    by "the analysis of 12 years of full
    disk Doppler velocity observations" LaBonte and Howard, (1982a),
    Howard and LaBonte, (1980);
      are present in the magnetic field rotation rate as well
      Snodgrass,  (1985,    1987b);
      Gilman and Howard, (1984;)
      Makarov et al., (1997)
    up to high latitudes as it is seen on the bottom plot of Fig. 1.
    The 11-year variability of the deviations of the period from the mean one
    in the sub-polar zones correspond to the torsional waves.
    The rotational rate of the pre-equatorial zones
    varies in time with a periodicity of 55--60 CR about
    (4 -- 5 years) Gavryuseva and Godoli, (2006).
    Results obtained by both methods  entirely agree with each other.

    This is better illustrated by Fig. 3
    where the correlation between the
    deviations at the different latitudes
    in the northern and in the southern hemispheres
    from the mean rotational rate
    corresponding to each latitude is plotted.
    The 11-year variability of the rotational rate at high latitudes
    is synchronized in both the hemispheres. The 5-year periodicity is
    common for pre-equatorial zones and active belts
    Gavryuseva and Godoli, (2006).

    This result confirms the increase of the rotational rate
    in those latitudinal zones where, in in some
    intervals of time, the magnetic field is weak.
    This result agrees with the conclusions of
      Stenflo  (1977, 1990);
      Makarov et al. (1997)  and
      Obridko and Shelting  (2001)
      that the pre-equatorial SMF rotation rate
      depends on the phase of the solar cycle
      and is higher during minimum of the activity.

      Interesting investigations have been performed 
      using helioseismological approuch to study how  
      rotation varies with radius and latitude within the solar interior. 
    Frequency splittings of acoustic modes confirm that the variation
    of rotation rate with latitude seen at the
    surface carries through the convection zone.
    At the base of the convective envelope (outer 30\% by radius) 
    there is the tachocline zone.
    R. Howe et al. (2000) "have detected changes in the rotation
    of the Sun near the base of its convective envelope, including
    a prominent variation with a period of 1.3 years at low latitudes.
 "Inversion of the global-mode frequency splittings reveals that 
 the largest temporal changes in the angular velocity are of the order
 of 6 nanohertz
 and occur above and below the tachocline that separates the Sun's
 differentially rotating convection zone  from the
 nearly uniformly rotating deeper radiative interior beneath". 
 This result agrees with the revealed variability of the SMF 
 at low latitudes.
 S. V. Vorontsov et al. (2002) confirm the presense of the bands
 of slower and faster rotation observed at the Sun's surface  and 
 migrated in latitude over the 11-year solar cycle. 
 The entire solar convective envelope appears to be involved in the
 torsional oscillations, with phase propagating poleward and equatorward
 from midlatitudes at all depths throughout the convective envelope.
 
 
  \section{Summary}
 \begin{enumerate}

 \item   
    The differential rotation of the large scale magnetic field
   and its temporal dependence have been investigated.
   The WSO data appear to indicate that the differential rotation
   of the magnetic field differs from that of the plasma at
   latitudes  above 55 degrees in both hemispheres.
 \item
   The 11-year periodicity
   of the rotational rate has been found at high latitudes.
  An 11-year periodicity  and a $5-$year quasi-periodicity
  of the rotational rate have been found
  in the near-equatorial zones and in the activity belts.
 
 \end{enumerate}
  
  This magnetic field topology, highly-organized over the solar surface
  and over time,
  must  be considered as a basic structure with
  a major influence  on the solar corona and solar wind propagation,
  and is fundamental for the understanding
  of the heliospheric structure and
  for the prediction of the magnetospheric perturbations.

\section*{Acknowledgments}
   I thank very much the WSO team for their great efforts
   in the measurements of the photospheric field.
   Thanks a lot to Prof. L. Paterno and Dr. E. Tikhomolov
   for precise and profitable advises and
   help in preparation of this paper.
   I am very grateful to Prof. B.T. Draine for his help
   in the revision  of this paper.


 \newpage
 \clearpage
 
 \begin{figure}
\centerline{
\includegraphics[angle=90, width=39pc]
   {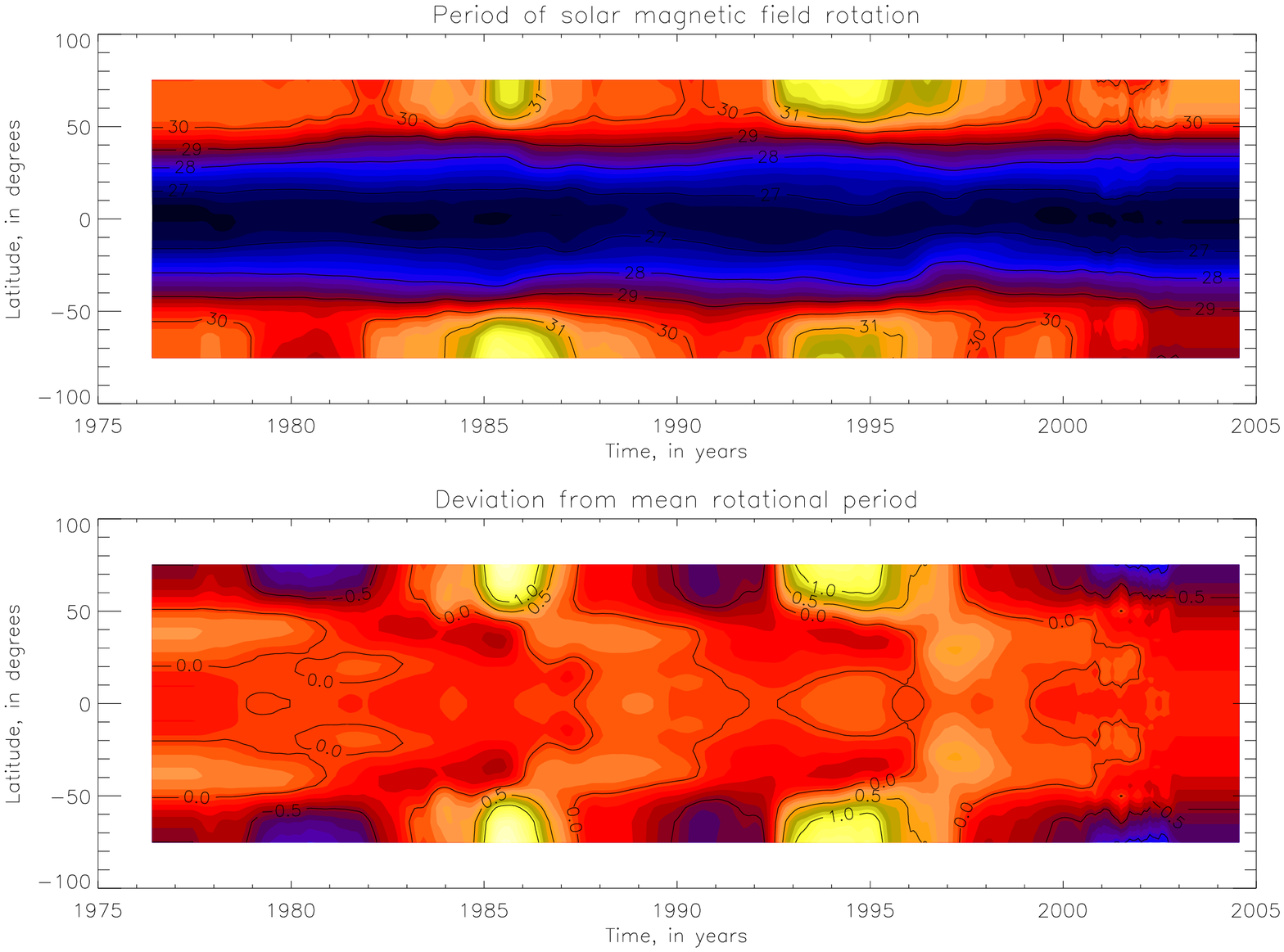}
}
   \caption
   {
    On the upper plot the sideral period of the  magnetic field
  differential rotation calculated by auto-correlation method for
  subsets of 3-year long
  is shown  as a function of time and latitude.
  Blue (yellow) colors correspond to the  shorter (longer) periods.
  Contours correspond to the sideral rotational periods of 
  27, 28, 29, 30 and 31  days.
  On the bottom plot the mean North-South deviation
  of the time dependent  rotational period from
  the differential rotational period averaged over 22 years
  is plotted  as a function of time and latitude.
  Blue (yellow) colors correspond to negative (positive) deviations.
  Contours correspond to the deviations of $0, \pm0.5, \pm1.0$  days.
 }
\end{figure}
 \begin{figure}
\centerline{
\includegraphics[angle=90, width=39pc]
   {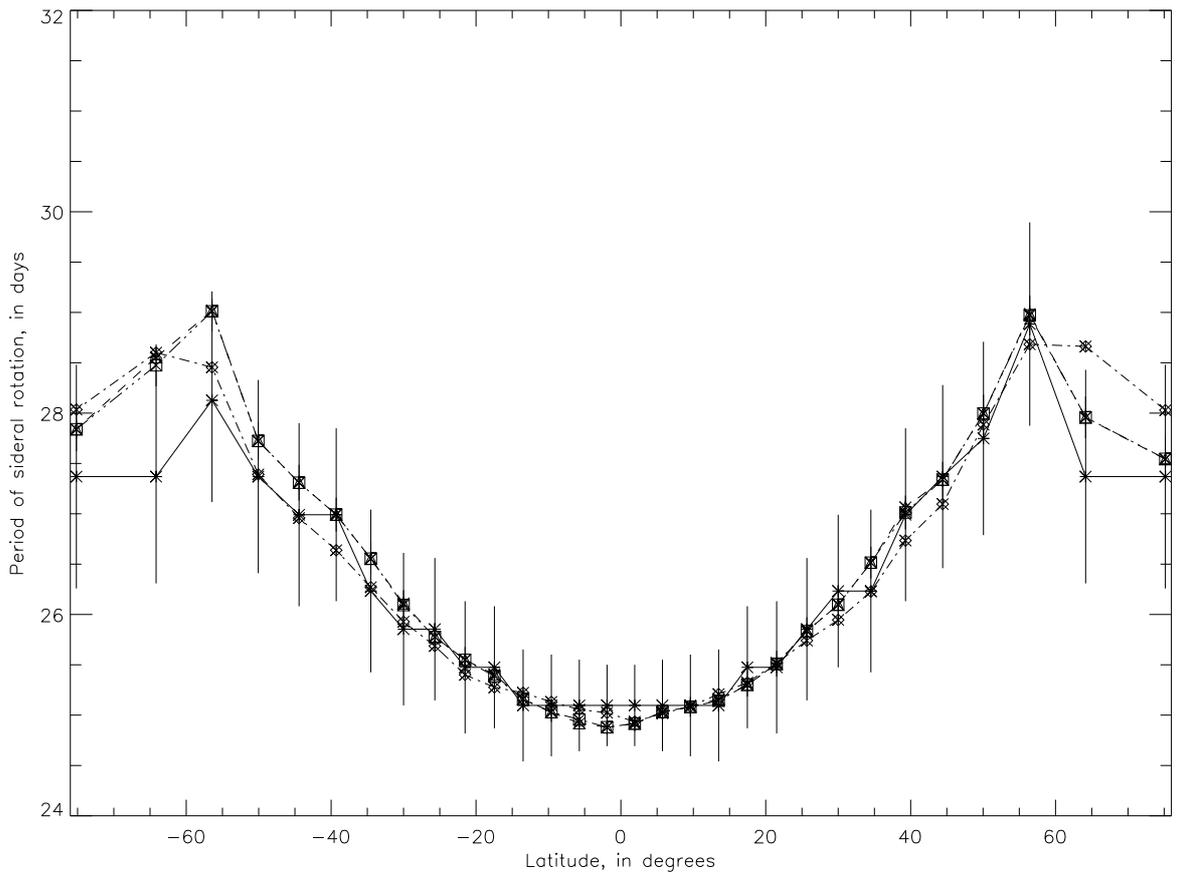}
}
   \caption   
 {  
The SMF mean rotational periods deduced from the full sets
are plotted as a function of latitude as it is deduced by auto-correlations method
 (continuous line) and by FFT method (broken line) and as a mean for sub-sets
 with corresponding error bars (dashed-dotted line).
 }
  \end{figure}
 \begin{figure}
\centerline{
\includegraphics[angle=90, width=39pc]
   {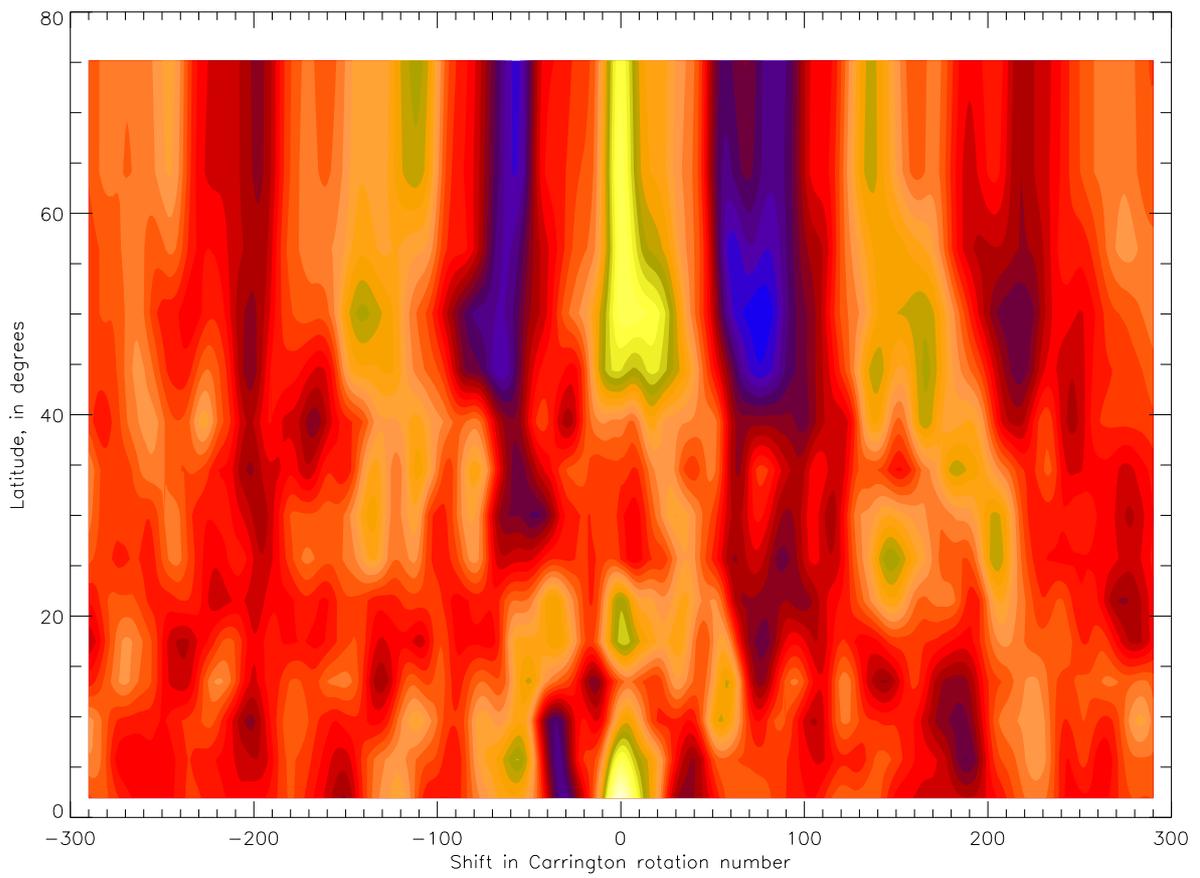}
}
\caption{
  Correlation between the deviation of the rotational period
  from the mean one  for each latitude
  on the northern and on the southern hemispheres
  as a function of time shift expressed in Carrington rotation number.
  }
  \end{figure}

   \end{document}